\documentclass[nofootinbib,twocolumn, prd, preprintnumbers]{revtex4-1}

\usepackage{amsmath}
\usepackage{graphicx}
\usepackage{dcolumn}
\usepackage{bm}
\usepackage[colorlinks=true, allcolors=blue]{hyperref}
\usepackage{xcolor}

\begin{document}


\title{Constraining neutrino charges at beam experiments}



\author{Jack D. Shergold}
\email{jack.d.shergold@ific.uv.es}
\affiliation{
 Instituto de Física Corpuscular (CSIC-Universitat de València),\\
Parc Científic UV, c/ Catedrático José Beltrán 2, E-46980, Paterna, Spain
}%

\author{Martin Spinrath}
\email{spinrath@phys.nthu.edu.tw}
\affiliation{%
 Department of Physics, National Tsing Hua University, Hsinchu 30013, Taiwan
}%
\affiliation{Physics Division, National Center for Theoretical Sciences, Taipei 10617, Taiwan}


\date{\today}

\begin{abstract}

We propose a new method to constrain neutrino charges at neutrino beam experiments. Uncharged in the Standard Model, evidence for a neutrino electric charge would be a smoking gun for new physics, shedding light on the Dirac or Majorana nature of neutrinos, and giving insight into the origin of charge quantization. We find that using the most sensitive magnetometers available, existing beam experiments could constrain neutrino charges $|q_{\nu}| \lesssim 10^{-13}$, in units of the electron charge, while future upgrades could strengthen these bounds significantly. We also discuss electromagnetic dipole moments and show that our proposal is highly sensitive to new long-range forces.
\end{abstract}

\maketitle


\section{Introduction}\label{sec:intro}

Due to their exclusively weak interactions and tiny masses, neutrinos remain the least understood fundamental particles in the Standard Model (SM). One important question is whether or not neutrinos couple to photons. In the Standard Model, where neutrinos are neutral and massless, there are no such couplings other than the charge radius, generated through radiative corrections~\cite{Giunti:2014ixa}. The observation of neutrino oscillations, however, confirms that neutrinos have mass, opening the possibility of a nonzero electric charge and electromagnetic dipole moments through the introduction of a right-chiral singlet neutrino. 

Among the electromagnetic moments, the neutrino charge is the most interesting, as a nonzero charge, however small, would call the origin of the apparent charge quantization observed among the other SM species into question, hinting at some softly broken symmetry at higher scales~\cite{Babu:1989tq, Babu:1989ex, Foot:1990uf, Foot:1992ui}. Constraints on the neutrino millicharge also provide key insight into the Dirac or Majorana nature of neutrinos. A nonzero, mass diagonal charge immediately signals that neutrinos are Dirac fermions. On the other hand, a set of flavor diagonal charges which sum exactly to zero strongly hints at neutrinos being Majorana fermions~\cite{Giunti:2024gec}. As such, any method which directly constrains neutrino-photon couplings is of great interest, not only to gain insight into the properties of neutrinos, but also to test key consequences of fundamental symmetries, such as charge conservation and the net neutrality of the universe in a model-independent way. For a general review on electromagnetic couplings of the neutrino, see, for instance, ~\cite{Giunti:2014ixa, Giunti:2024gec}.

In this paper we propose a method to constrain the electric charge of neutrinos, $q_\nu e$, using beam experiments, which could, in principle, be extended to the dipole moments. Dense, ultrarelativistic bunches of charged neutrinos act as substantial current sources, giving rise to large azimuthal magnetic fields which could be picked up by a sensitive magnetometer placed alongside the beam trajectory. When combined with details of the bunch structure and composition, measurements of these magnetic fields yield stringent constraints on the individual neutrino charges.

The current model-dependent upper bound on the mass-diagonal charges from cosmology is $|q_{\nu_i}|\lesssim 10^{-35}$~\cite{Caprini:2003gz,ParticleDataGroup:2024cfk}. Laboratory and astrophysical bounds are typically much weaker. The strongest astrophysical bounds on the flavor-diagonal charges, $|q_{\nu_\alpha}| \lesssim 10^{-19}$, come from the neutron star turning mechanism~\cite{Studenikin:2012vi}. If charge conservation in $\beta$-decays and the net neutrality of matter are assumed, the strongest laboratory bound on the mass-diagonal charge is then, $|q_{\nu_i}| \simeq 10^{-21}$~\cite{Giunti:2014ixa}, while the most stringent
model-independent laboratory bounds on the individual flavor-diagonal charges is $|q_{\nu_\alpha}| \lesssim 10^{-13}$~\cite{AtzoriCorona:2022jeb, A:2022acy, Giunti:2023yha}. Our proposal is model-independent for neutrino couplings to photons, and can be extended to new long-range forces with minimal assumptions.

The remainder of this paper will be structured as follows. In Sec.~\ref{sec:nu_fields} we will introduce our method for constraining neutrino charges at beam experiments, and discuss its prospects at existing and future neutrino beams in Sec.~\ref{sec:exp}. We will then comment on the possibility of constraining other electromagnetic moments of the neutrino and new long-range forces with this method in Sec.~\ref{sec:prospects}, before concluding in Sec.~\ref{sec:conclusions}. 
\section{Neutrino magnetic fields}\label{sec:nu_fields}
A single charged neutrino at rest with charge $q_\nu e$ generates an electric field
\begin{equation}
\vec{E}_{\nu,\mathrm{rest}} = \frac{q_\nu e}{4 \pi \epsilon_0} \frac{\hat{r}_0}{r_0^2},
\end{equation}
where $r_0$ and $\hat{r}_0$ are the rest frame distance from the neutrino and the unit vector pointing away from it, respectively, $c$ is the speed of light, and $\epsilon_0$ is the permittivity of free space. The same neutrino, as seen in the frame moving with relative velocity $-\vec{v}_\nu$ generates a magnetic field, see, e.g.,~\cite{Jackson:1998nia},
\begin{equation}\label{eq:1nufield}
    \vec{B}_\nu =  \frac{q_\nu e}{4 \pi \epsilon_0 c}  \frac{\gamma_\nu|\vec{\beta}_\nu| R_\perp \hat{e}_\phi}{\left(R_\perp^2 + \gamma_\nu^2 R_\parallel^2\right)^{\frac{3}{2}}},
\end{equation}
where $\gamma_\nu$ is the neutrino Lorentz factor, $\vec{\beta}_\nu = \vec{v}_\nu/c$, $\hat{e}_\phi$ is the azimuthal unit vector, anticlockwise about the beam direction $\hat{\beta}_{\nu}$, and $R_\parallel$ and $R_\perp$ are the distances between  the observer and the neutrino in the moving frame coordinates, parallel and perpendicular to the neutrino direction of flight, respectively. An observer exactly perpendicular to an ultrarelativistic neutrino therefore experiences a maximum magnetic field with magnitude
\begin{equation}
    B_{\nu,\mathrm{max}} = \frac{q_\nu e}{4 \pi \epsilon_0 c}  \frac{|\vec{\beta}_\nu| \gamma_\nu}{R_\perp^2}.
\end{equation}
That is, a magnetic field enhanced by an enormous factor $\gamma_\nu$ relative to a nonrelativistic neutrino. This remarkable result is precisely what we expect at a neutrino beam experiment such as LBNF~\cite{DUNE:2015lol} or J-PARC~\cite{JPARC}, where neutrinos have energy $E_{\nu} \sim \mathcal{O}(\mathrm{GeV})$, and correspondingly a Lorentz factor
\begin{equation}
    \gamma_{\nu} = \frac{E_\nu}{m_{\nu} c^2} = 10^{12} \left[\frac{E_\nu}{1\,\mathrm{GeV}}\right]\left[\frac{1\,\mathrm{meV}}{m_\nu c^2}\right],
\end{equation}
leading to sizable transverse magnetic fields even for millicharged neutrinos. 

This is the magnetic field generated by a single neutrino. A neutrino beam experiment will instead feature huge bunches of neutrinos with some spatial and energy distribution, $f(\vec{p}_{\nu}, \vec{r}_{\nu})$, whose individual magnetic fields sum to the total magnetic field
\begin{equation}\label{eq:Bint}
    \vec{B}_\nu =  \frac{q_\nu e}{4 \pi \epsilon_0 c}  \int \mathrm{d}^3 p_\nu\, \mathrm{d}^3 r_\nu \, f(\vec{p}_{\nu}, \vec{r}_\nu) \frac{\gamma_\nu|\vec{\beta}_\nu| r_\perp  \hat{e}_\varphi
    }{\left(r_{\perp}^2 + \gamma_\nu^2 r_{\parallel}^2\right)^{\frac{3}{2}}},
\end{equation}
for an observer at the origin, where we have assumed that all neutrinos fly along the same direction, $\vec{p}_\nu$ and $\vec{r}_{\nu}$ are the neutrino momentum and position, respectively, and all quantities are in the moving or \textit{laboratory} frame. Note that due to the net motion of the neutrino bunch, $\vec{B}_{\nu}$ will be a function of time. We will comment on the consequences of this for our proposal shortly. Finally, the neutrino distribution function is normalized to $N_\nu$, the number of neutrinos in each bunch which is given in terms of the proton beam power, $P_\mathrm{beam}$, pulse rate, $\omega$, and proton energy by
\begin{equation}
    \frac{N_{\nu}}{\varepsilon} = 6 \cdot 10^{11}\, \left[\frac{P_\mathrm{beam}}{1\,\mathrm{MW}}\right]\left[\frac{100\,\mathrm{GeV}}{E_p}\right]\left[\frac{1\,\mathrm{Hz}}{\omega}\right] \left[\frac{100}{N_b}\right],
\end{equation}
with $\varepsilon \sim \mathcal{O}(1)$ the number of neutrinos produced per proton, and $N_b$ the number of bunches per pulse.

To illustrate the power of this technique for constraining neutrino charges, let us first focus on the simplest case of a densely packed, monochromatic neutrino beam. This corresponds to the distribution function
\begin{equation}\label{eq:denseBunch}
    f(\vec{p}_{\nu}, \vec{r}_\nu) = N_\nu \delta^{(3)}(\vec{p}_\nu - \vec{k}_\nu) \delta^{(3)}(\vec{r}_\nu - \vec{R}_{\nu}),
\end{equation}
with $\vec{R}_\nu$ the time-dependent vector from the neutrino to the observer, and $\vec{k}_\nu$ the fixed momentum of the bunch. This leads to a maximum magnetic field directly perpendicular to the ultradense bunch

\begin{align}\label{eq:perfectBeam}
        \frac{B_{\nu,\mathrm{max}}}{|q_\nu|} &=  N_\nu \frac{q_\nu e}{4 \pi \epsilon_0 c}  \frac{\gamma_\nu|\vec{\beta}_\nu| }{R_\perp^2} \notag \\
        &= 2\cdot 10^{9}\,\mathrm{T} \,\left[\frac{N_\nu}{10^{12}}\right] \left[\frac{\gamma_\nu}{10^{12}}\right] \left[\frac{50\,\mathrm{mm}}{R_\perp}\right]^2.
\end{align}
The most sensitive SQUID magnetometers are sensitive to magnetic fields as small as $|\vec{B}_{\nu}| \simeq 10^{-15}\,\mathrm{T}$~\cite{JacksonKimball:2017elr}, implying a sensitivity to neutrino charges as small as $|q_{\nu}| \sim \mathcal{O}(10^{-24})$, far beyond existing laboratory and astrophysical constraints. Alternatively, a SERF magnetometer sensitive to $|\vec{B}_{\nu}| \simeq 10^{-17}\,\mathrm{T}$ could push this bound even further~\cite{Kominis:2003gax}.

However, there are two issues with this claim. The first relies on the magnetometer being able to measure the almost instantaneous pulse of maximum magnetic field when $R_\parallel = 0$, outside of which there is almost no magnetic field at the detector site owing to the $1/\gamma_\nu^3$ suppression. A rough estimate for this time period is $\Delta t \simeq R_\perp/(\gamma_\nu c) \sim \mathcal{O}(10^{-22}\,\mathrm{s})$ for the choice of parameters given in~\eqref{eq:perfectBeam}, necessitating magnetometers with time resolution far beyond what is currently achievable. The second issue is in our choice of distribution function, $f(\vec{p}_{\nu}, \vec{r}_\nu)$, in~\eqref{eq:denseBunch}, in particular in our assumption of an ultradense bunch. In practice, each bunch will have some finite size along the direction of flight, of which only a small fraction within the window $c \Delta t$ will contribute to the magnetic field, provided that the bunch size exceeds the window size. As we will see, properly accounting for this effect will lead to an effective cancellation of the enhancement factor, $\gamma_\nu$, in realistic examples but will in turn allow us to relax the temporal resolution requirements of the dense bunch.

\section{Realistic example}\label{sec:exp}

We now turn our attention to the more realistic case of a finite bunch size. A neutrino beam is generated by a primary proton beam split into bunches with interval $\tau \sim \mathcal{O}(\mathrm{ns})$, such that the characteristic length scale of the bunch along the direction of flight is $l_\nu = c\tau \sim \mathcal{O}(\mathrm{m})$.  We will neglect the finite size of the bunch in the transverse direction, as the opening angle of the neutrino beam at production will be heavily suppressed by a factor $1/\gamma_\nu$. 

Rather than a constant line density, we will consider Gaussian bunches, where the beam is distributed normally along the direction of flight, with standard deviation $\sigma_\nu = l_\nu$. This is a good approximation for simple beams. When more sophisticated slicing procedures are applied, however, the proton bunches can have a more complicated geometry~\cite{Ganguly:2024lqh}.

In this scenario, the distribution function of the neutrino bunch is found by making the replacement
\begin{equation}\label{eq:gaussianBunch}
    \delta(r_\parallel - R_\parallel) \to  \frac{1}{\sigma_\nu \sqrt{2\pi}} \exp\left(-\frac{1}{2}\left[\frac{r_{\parallel}-R_\parallel}{\sigma_\nu}\right]^2\right)
\end{equation}
in~\eqref{eq:denseBunch}, with $R_{\parallel}$ evolving in time as the neutrino bunch passes the detector. For simplicity, we treat the beam as monochromatic, as the spread in neutrino energies makes little difference provided that they are all roughly the same order of magnitude. We demonstrate this explicitly in Appendix~\ref{app:time_resolution}.

Integrating over the Gaussian bunch using~\eqref{eq:Bint}, we arrive at the maximum magnetic field generated by the bunch
\begin{equation}
    \label{eq:Bnu_qnu_approx}
    B_{\nu,\max} = N_\nu \frac{\, q_\nu \, e}{(2\pi)^\frac{3}{2} \epsilon_0 c} \frac{|\vec{\beta}_\nu|}{R_\perp \sigma_\nu} + \mathcal{O}\left(\frac{R_\perp}{\gamma_\nu \sigma_\nu}\right),
\end{equation}
which persists for a duration $\Delta t \simeq \tau$. As such, we have gained the ability to resolve our signal in time, at the expense of the huge $\gamma_\nu$ enhancement of the magnetic field occurring for ultrarelativistic, pointlike bunches. This demonstrates an important property. By constricting the beam along the direction of flight, we can increase the magnetic field generated by the beam, provided that we also improve the magnetometer time resolution by a similar factor. Alternatively, we can use more diffuse beams to compensate for magnetometers with bad time resolution. Finally, in the extreme limit $ \sigma_\nu \ll R_\perp/\gamma_\nu$, we recover the $\gamma_\nu$ scaling of the ultradense bunch.

As before, we can also estimate the magnitude of this magnetic field
\begin{equation}\label{eq:gaussianBeam}
    \frac{B_{\nu,\mathrm{max}}}{|q_\nu|} = 2\cdot 10^{-4}\,\mathrm{T} \,\left[\frac{N_\nu}{10^{12}}\right] \left[\frac{1\,\mathrm{ns}}{\tau}\right]\left[\frac{50\,\mathrm{mm}}{R_\perp}\right],
\end{equation}
which, as expected, is significantly smaller than the ultradense bunch estimate~\eqref{eq:perfectBeam}. Nevertheless, a SQUID magnetometer could still set bounds on the neutrino charge $|q_\nu| \sim \mathcal{O}(10^{-11})$, competitive with existing model-independent bounds on the neutrino charge. We additionally note that, unlike the ultradense bunch, this bound is independent of the neutrino mass to leading order due to the absence of the Lorentz factor in~\eqref{eq:gaussianBeam}. 

We tabulate the experimental parameters for a variety of existing and upcoming neutrino beam experiments, alongside the maximum magnetic field at $R_\perp = 50\,\mathrm{mm}$, assuming Gaussian bunches with distribution function~\eqref{eq:gaussianBunch} in Table~\ref{tab:expParams}. Of the beams considered, an experiment at J-PARC currently offers the greatest sensitivity to neutrino charges, $|q_{\nu}| \sim \mathcal{O}(10^{-11})$ with a SQUID magnetometer, or $|q_\nu| \sim \mathcal{O}(10^{-13})$ with a SERF magnetometer. Looking to the future, upgrades to LBNF (denoted herein as LBNF-u), including doubling the beam power and lowering the bunch interval to $\tau \simeq 0.07\,\mathrm{ns}$~\cite{DUNE:2016evb,Ganguly:2024lqh,Angelico:2019gyi}, could see an improvement over J-PARC by a factor $\sim 3$, pushing the sensitivity to the neutrino charge toward $|q_\nu| \sim \mathcal{O}(10^{-14})$.

We note that one could also measure electric fields due to neutrino charges in a similar manner. However, magnetometers are typically far more sensitive than electrometers. For instance, the electrometer discussed in~\cite{Facon:2016ary} is sensitive to electric fields with magnitude $1.2\,\mathrm{mV}\,\mathrm{cm}^{-1}$, corresponding to the much weaker bound $|q_\nu| \lesssim 10^{-6}$ at J-PARC.

\begin{table}
    \renewcommand{\arraystretch}{1.3}
    \begin{ruledtabular}
    \begin{tabular}{ccccccc}
        Beam & $N_\nu$ & $\tau\,(\mathrm{ns})$ & $E_\nu\,(\mathrm{GeV})$ & $B_{\nu,\mathrm{max}}/|q_{\nu}|\,(\mathrm{T})$ \\
        \hline
        LBNF-u & $3.0\cdot 10^{11}$ & 0.07 & 2.5 & $1.1 \cdot 10^{-3}$\\
        J-PARC & $4.0\cdot 10^{13}$ & 27 & 0.6 & $3.8 \cdot 10^{-4}$\\
        LANSCE & $3.1 \cdot 10^{13}$ & 93 & 0.05 & $8.5 \cdot 10^{-5}$\\
        ESSnuSB (ESS) & $2.8\cdot 10^{14}$ & 1300 & 0.3 & $5.5 \cdot 10^{-5}$\\
        LBNF & $1.5\cdot 10^{11}$ & 1 & 2.5 & $3.8\cdot 10^{-5}$\\
        nuSTORM & $1.9\cdot 10^{10}$ & 2 & 2.0 & $2.4 \cdot 10^{-6}$\\
        ENUBET & $9.4 \cdot 10^{11}$ & $3\cdot 10^6$ & 3.1 & $8.0 \cdot 10^{-11}$\\
    \end{tabular}
    \end{ruledtabular}
    \caption{Experimental parameters and estimated magnetic fields at $R_\perp = 50\,\mathrm{mm}$ for a variety of neutrino beam experiments. Parameters taken from~\cite{Friend:2019fuq, CCM:2021leg, ESSnuSB:2023ogw, Garoby:2017vew, DUNE:2016evb, Ganguly:2024lqh, Angelico:2019gyi, Ahdida:2020whw, Longhin:2714046, ENUBET:2023hgu}, assuming $\varepsilon = 1$ and a Gaussian bunch in all cases. LBNF-u denotes the proposed upgrades to the LBNF beam.
    }
    \label{tab:expParams}
\end{table}

\begin{figure*}
    \centering
    \includegraphics[width=\linewidth]{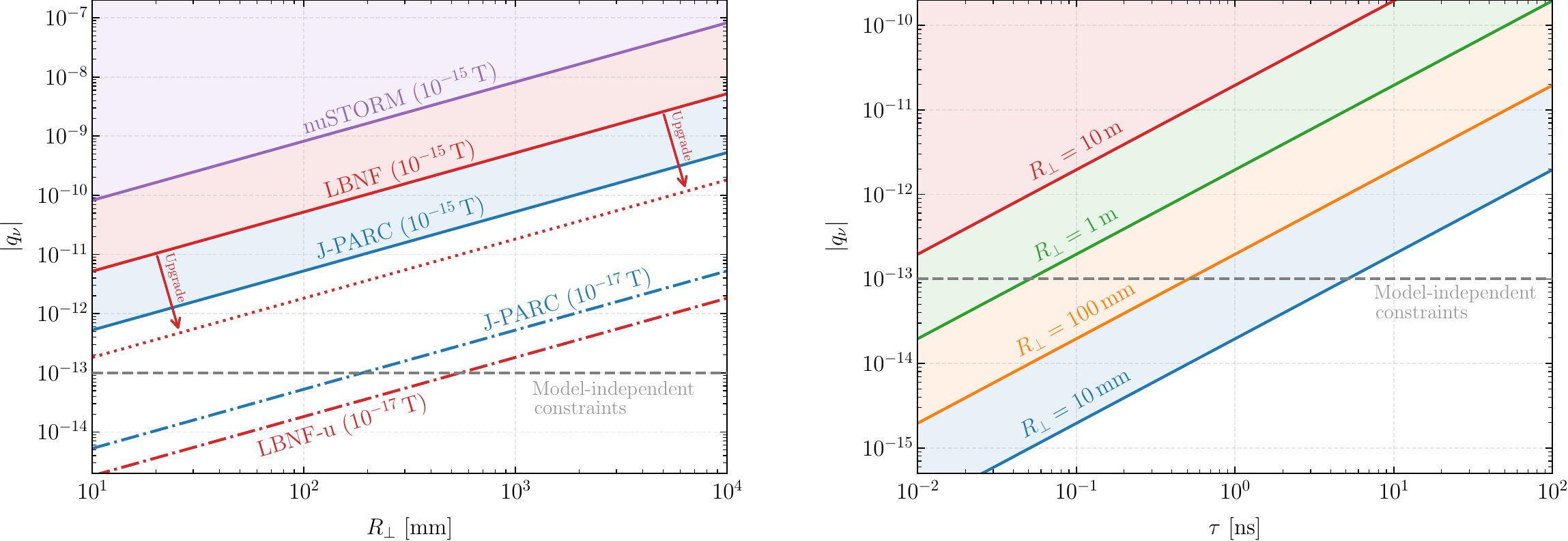}
    \caption{Sensitivity to the neutrino charge for left panel: the most promising existing and upcoming neutrino experiments, cf. Table~\ref{tab:expParams}, and right panel: a toy experiment with variable bunch size, but otherwise the same beam parameters as J-PARC. In both panels, the solid lines assume a magnetometer with sensitivity comparable to existing SQUIDs, $B_\mathrm{ref} \simeq 10^{-15}\,\mathrm{T}$, while the dot-dashed lines assume sensitivity to $B_\mathrm{ref}\simeq 10^{-17}\,\mathrm{T}$, akin to a SERF magnetometer. The horizontal dashed line represents the current, most stringent, model-independent laboratory bounds on the individual flavor-diagonal charges, $|q_{\nu_\alpha}| \lesssim 
    10^{-13}$~\cite{AtzoriCorona:2022jeb, A:2022acy, Giunti:2023yha}.}
    \label{fig:qnu_sensitivity_combined}
\end{figure*}

In Fig.~\ref{fig:qnu_sensitivity_combined} we show the sensitivity to $|q_\nu|$ as a function of the transverse distance to the beam, $R_\perp$, and the bunch spacing, $\tau = \sigma_\nu/c$. In all cases, we use the full result found by solving~\eqref{eq:Bint} for a Gaussian bunch, but note that the expansion~\eqref{eq:Bnu_qnu_approx} agrees to extreme precision. Thus we recover both the approximate $1/R_\perp$ and $1/\tau$ dependence of the induced magnetic field in both figures.

In the left panel of Fig.~\ref{fig:qnu_sensitivity_combined}, we focus on current and upcoming neutrino beams, and see that with current SQUID technology it would be difficult to overcome existing constraints. However, with future upgrades to neutrino beams, along with more sensitive magnetometers one might be able to compete with existing model-independent laboratory constraints.

Seemingly, the most promising way to improve the sensitivity of such an experiment is to improve the neutrino density, which can be achieved by either increasing the beam power or by reducing the bunch size. Reducing the bunch size comes with the added benefit of increased control over flux and cross-section systematic uncertainties at oscillation experiments~\cite{Ganguly:2024lqh}, and so is a natural direction of improvement.

Before moving on, we would like to comment on the time resolution requirements of the magnetometer and potential backgrounds. If the time resolution of the magnetometer is insufficient to resolve a single bunch, it may still be able to set competitive constraints by averaging over several bunches within a single beam spill, typically lasting $\mathcal{O}(\mu\mathrm{s})$, in comparison to the $\mathcal{O}(\mathrm{ns})$ interval of a single bunch. The trade off is that the neutrino current density will by diluted by the interbunch spacing, resulting in bounds on the neutrino charge $\sim \mathcal{O}(10)$ weaker for the experiments considered here. This is discussed at length in Appendix~\ref{app:time_resolution}.

Regarding backgrounds, charged particles from cosmic rays, the beam source, or natural radioactivity could in theory spoil any attempt at measuring the neutrino magnetic field. Any moving charge will create a magnetic field. However, many of those backgrounds can be significantly reduced or distinguished from our signal. First of all, provided that the detector is far enough down the neutrino beamline, any spillage of charged particles from the source will be exponentially suppressed by material between the source and detector. Similarly, placing the detector underground will shield from the majority of cosmic rays.

There are also some intrinsic properties of the background which will not be shared by many potential backgrounds, in particular the directionality and timing information. A magnetometer with sufficient time resolution will be able to trigger precisely on the windows where a neutrino bunch is passing; any field measured outside of these windows cannot be due to neutrinos, and can safely be discarded. Even if the resolution is insufficient to resolve a single bunch, we expect that the field due to other charged particles will be significantly larger than that of neutrinos, and have an irregular time structure, and so can easily be distinguished. Magnetic fields which are not orthogonal to the neutrino beamline can also be discarded.
SQUID loops are very small devices, as small as a few hundred nanometers \cite{Aprili:2006}, that measure the magnetic flux in a given direction, and as such can be oriented to be most sensitive in the expected signal direction. A SERF on the other hand tends to be larger, on the order of cm$^3$, but creates a three dimensional map of the magnetic field. As such, one could direct the beam through a SERF, and compare the magnitude and directionality with the expectation from the beam.

\section{Additional prospects}\label{sec:prospects}

It is also possible to constrain neutrino properties other than the neutrino charge using this technique. In this section we will comment on constraining electromagnetic moments other than the neutrino charge, and new long-range forces.

\subsection{Other electromagnetic moments}\label{sec:moments}

We begin by briefly discussing the prospects of such an experiment for constraining other neutrino electromagnetic moments. In particular, we will focus on the neutrino magnetic dipole (MDM) and electric dipole (EDM) moments. 

An ultrarelativistic neutrino traveling along the direction $\hat{\beta}_\nu$ with helicity $h_\nu = \pm 1$ will give rise to magnetic fields
\begin{equation}
    \vec{B}_{\nu}^{\mu} = \frac{\mu_\nu h_\nu}{8\pi \epsilon_0 c^2}\frac{\gamma_\nu^2  R_\parallel R_\perp \hat{e}_\rho}{\left(R_\perp^2 + \gamma_\nu^2 R_\parallel^2\right)^\frac{5}{2}},
\end{equation}
due to the intrinsic MDM, $\mu_\nu$, of the neutrino, where $\hat{e}_\rho$ is the unit vector pointing outward from the beam axis. The corresponding magnetic field of the EDM can be found by replacing $\mu_\nu \hat{e}_\rho \to 3|\vec{v}_\nu| \varepsilon_\nu \hat{e}_\varphi$, with $\varepsilon_\nu$ the intrinsic EDM of the neutrino. We neglect the field along the beam axis for the MDM, which does not receive a $\gamma_\nu$ enhancement. 

Unlike the field due to the electric charge, which has a maximum at the instant when $R_\parallel = 0$, the field due to the dipole moments has two maxima when $R_\parallel = R_\perp/(2\gamma_\nu)$. However, as these maxima are separated in time by $\Delta t_\mathrm{peak} \simeq R_\perp/(\gamma_\nu c) \sim \mathcal{O}(10^{-22}\,\mathrm{s})$, they are unresolvable for all practical purposes. Taking into account the finite width of each bunch, assuming the Gaussian bunch~\eqref{eq:gaussianBunch}, we find the maximum magnetic fields due to the MDM
\begin{align}
    \frac{B_{\nu,\mathrm{max}}^{\mu}}{\widetilde{\mu}_\nu} &= N_\nu \frac{\mu_0}{6 (2\pi)^{\frac{3}{2}}} \frac{1}{R_\perp^2 \sigma_\nu} + \mathcal{O}\left(\frac{R_\perp}{\gamma_\nu \sigma_\nu}\right) \notag \\
    &= 2\cdot 10^{-16}\,\mathrm{T}\,\left[\frac{N_\nu}{10^{12}}\right]\left[\frac{1\,\mathrm{ns}}{\tau}\right]\left[\frac{50\,\mathrm{mm}}{R_\perp}\right]^2,
\end{align}
in terms of the normalized MDM, $\widetilde{\mu}_\nu = \mu_\nu/\mu_B$, where $\mu_B$ is the Bohr magneton. The bound on the normalized EDM, $\widetilde{\varepsilon}_\nu c = \varepsilon_\nu c/\mu_B$, is approximately three times stronger. As with the field due to the electric charge, both persist for a duration $\Delta t \simeq \tau$. Unfortunately, even for the most optimistic magnetometer with sensitivity to magnetic fields $|\vec{B}_\nu| \simeq 10^{-17}\,\mathrm{T}$, these bounds are nowhere near competitive with existing constraints on the dipole moments, which constrain the combination $|\lambda_\nu| = |\mu_\nu - i\varepsilon_\nu c| \lesssim 10^{-12} \mu_B$~\cite{Giunti:2024gec}.
\subsection{Additional long range forces}\label{sec:newU1}
In addition to an electric charge, it is entirely possible that the neutrino partakes in interactions involving some new long-range force. For example, the neutrino could carry a charge under some new U(1) gauge group, or have long-range interactions with other SM fermions induced by the kinetic mixing of a new U(1) gauge boson with those of the SM. Well motivated, anomaly-free examples of such groups include $B-L$, with $B$ and $L$ the baryon and lepton number, respectively, and combinations of $L_\alpha - L_\beta$, with $\alpha,\beta \in \{e,\mu,\tau\}$. A comprehensive discussion of hidden photon models can be found in e.g.~\cite{Bauer:2018onh,Cline:2024qzv,Foldenauer:2019vgn}. Other possibilities include, but are not limited to, light scalar exchange, or fermion pair exchange interactions.

For the purposes of our discussion, we will assume a Yukawa-like potential, $\phi_\nu$, due to the exchange of a single particle, which gives rise to the field $\vec{D}_\nu = - \vec{\nabla} \phi_\nu$ in the rest frame of the neutrino, analogous to the electric field in classical electromagnetism. In the same way as we did in Sec.~\ref{sec:nu_fields}, we can then transform into the frame moving at velocity $-\vec{v}_{\nu}$ with respect to the neutrino rest frame to find the analogy of the magnetic field due to the new interaction
\begin{widetext}
    \begin{equation}
    \vec{C}_\nu = \frac{\mathcal{Q}_\nu e}{4 \pi \epsilon_0 c}  \left[ \frac{\gamma_\nu|\vec{\beta}_\nu| R_\perp \hat{e}_\varphi}{(R_\perp^2 + \gamma_\nu^2 R_\parallel^2)^\frac{3}{2}} + \frac{\gamma_\nu|\vec{\beta}_\nu| R_\perp \hat{e}_\varphi}{\lambda_X (R_\perp^2 + \gamma_\nu^2 R_\parallel^2)} \right] \exp\left[ - (R_\perp^2 + \gamma_\nu^2 R_\parallel^2)^\frac{1}{2}/\lambda_X \right],
\end{equation}
\end{widetext}
where $\lambda_X = \hbar/(M_X c)$ is the effective range of the new force, given in terms of the mediator mass, $M_X$, and $\mathcal{Q}_\nu$ is the coupling of neutrinos to the new mediator, normalized to the elementary charge. It should be immediately clear that this field will be heavily suppressed if $\lambda_X \gg R_\perp$, such that we are only able to constrain new forces with mediators of mass
\begin{equation}\label{eq:massBounds}
    M_X \lesssim 10^{-8}\,\frac{\mathrm{eV}}{c^2}\,\left[\frac{R_\perp}{50\,\mathrm{mm}}\right],
\end{equation}
in which case $\vec{C}_{\nu}$ behaves to a very good approximation as $\vec{B}_\nu$ under the transformation $q_\nu \to \mathcal{Q}_\nu$. However, it is not as simple as making the same replacement in~\eqref{eq:lineBeam} to derive a constraint on $\mathcal{Q}_\nu$, as we must also consider the detector response to $\vec{C}_\nu$. This depends on the coupling of the detector particles to the mediator, $\mathcal{Q}_d e$, such that the \textit{effective} magnetic field felt by the detector is
\begin{equation}\label{eq:lineBeam}
    \frac{B_{\nu,\mathrm{max}}}{|\mathcal{Q}_\nu \mathcal{Q}_d|} = 2\cdot 10^{-4}\,\mathrm{T} \,\left[\frac{N_\nu}{10^{12}}\right] \left[\frac{1\,\mathrm{ns}}{\tau}\right] \left[\frac{50\,\mathrm{mm}}{R_\perp}\right],
\end{equation}
allowing us to constrain combinations of couplings $|\mathcal{Q}_{\nu} \mathcal{Q}_d| \sim \mathcal{O}(10^{-11})$ with a SQUID magnetometer, or $|\mathcal{Q}_{\nu} \mathcal{Q}_d| \sim \mathcal{O}(10^{-13})$ with a SERF magnetometer. The dominant constraints in the mass range~\eqref{eq:massBounds} for SM fermions charged under a new U(1), e.g. $B-L$, $L_\mu-L_e$, are given by fifth force and equivalence principle experiments, and are far stronger than what can be achieved using this technique. See~\cite{Wise:2018rnb,Adelberger:2009zz, Wagner:2012ui}, and references therein. Similarly, light scalar-induced electron-neutrino interactions are heavily constrained~\cite{Venzor:2020ova}. One possibility, however, is that of a dark photon which kinetically mixes with the SM gauge bosons. If the dark photon does not comprise the relic dark matter density, then our method is competitive with existing bounds, which are of order $|\mathcal{Q}_{\nu}| \sim  |\mathcal{Q}_d| \sim \mathcal{O}(10^{-6} - 10^{-7})$~\cite{Caputo:2020bdy, Garcia:2020qrp, Betz:2013dza, Jaeckel:2010xx, Fabbrichesi2021,Li:2023vpv}. It should be stressed that while we have focused on Yukawa-like potentials here, our technique is broadly applicable to long-range forces, and may yield competitive constraints for a wide variety of models.

\section{Conclusions}\label{sec:conclusions}
In this paper, we have proposed a novel method to constrain the electric charge of neutrinos using beam experiments. By placing a highly sensitive magnetometer along the path of a dense, ultrarelativistic neutrino bunch, we have demonstrated that it is possible to set constraints on the neutrino charge, competitive with existing model-independent laboratory searches. 

The efficacy of our proposal relies heavily on the neutrino beam structure and magnetometer sensitivity. Denser neutrino bunches have the potential to provide stronger constraints on the neutrino charge, but require magnetometers with better time resolution. We have shown, however, that this can be alleviated by averaging over multiple bunches, at a slight cost to sensitivity. Of the experiments considered, we found that J-PARC currently has the best potential to constrain the neutrino charge, but that future upgrades to the LBNF beam would allow for constraints as strong as $|q_\nu|\lesssim 10^{-14}$. 

We have also explored the possibility of constraining the magnetic and electric dipole moments of the neutrino, and the sensitivity to new long-range forces. while our potential constraint on dipole moments are uncompetitive with existing limits, our method shows promise for probing new long-range forces mediated by light bosons.

\begin{acknowledgments}
We are incredibly grateful to Gary Barker, John Back, Martin Bauer, and Valentina de Romeri for their helpful comments during the preparation of this manuscript. J. D. Shergold is supported by the Spanish Grants No. PID2023-147306NB-I00 and No. CEX2023-001292-S (MCIU/AEI/
10.13039/501100011033), as well as CIPROM/2021/054 (Generalitat Valenciana). M. S. is supported by the National Science and Technology Council (NSTC) of Taiwan under Grants No. NSTC 112-2112-M-007-039 and No. NSTC
114-2112-M-007-030.
\end{acknowledgments}

\appendix
\section{Imperfect time resolution and polychromatic beams}\label{app:time_resolution}
In this Appendix we will derive the effects of imperfect time resolution, and any spread in beam energy on the expected signals. We begin with~\eqref{eq:Bint}, now including the explicit time dependence of the neutrino distribution function
\begin{equation}\label{eq:Bint_t}
    \vec{B}_\nu(t) =  \frac{q_\nu e}{4 \pi \epsilon_0 c}  \int \mathrm{d}^3 p_\nu\, \mathrm{d}^3 r_\nu \, f(\vec{p}_{\nu}, \vec{r}_\nu, t) \frac{\gamma_\nu|\vec{\beta}_\nu| r_\perp  \hat{e}_\varphi
    }{\left(r_{\perp}^2 + \gamma_\nu^2 r_{\parallel}^2\right)^{\frac{3}{2}}}.
\end{equation}
We next decompose the distribution into its spatio-temporal and momentum components as
$f(\vec{p}_{\nu}, \vec{r}_\nu, t) = f_p(\vec{p}_\nu) f_r(\vec{r}_\nu,t)$, with $f_p$ some normalized, but unspecified momentum distribution. Since we are now focusing on multiple bunches, each arriving at different times, the spatial distribution function is a sum over single particle distributions, which we assume are Gaussian
\begin{align}
     f_r(\vec{r}_\nu,t) &= \frac{N_\nu}{\sigma_\nu \sqrt{2\pi}} \delta^{(2)}(\vec{r}_\perp - \vec{R}_\perp)\\ \notag
     &\qquad\times\sum_{n=-\infty}^\infty\exp\left(-\frac{1}{2}\left[\frac{r_{\parallel}-R_{\parallel,n}(t)}{\sigma_\nu}\right]^2\right), \\
     R_{\parallel,n}(t) &=\widetilde{R}_\parallel + c(t-t_0) + n\Delta 
\end{align}
with $\Delta \gg \sigma_\nu$ the interbunch distance, while $\widetilde{R}_\parallel$ and $t_0$ are some reference starting position and time, respectively, which we can set to zero. With these choices, the $n = 0$ bunch sits perfectly perpendicular to the detector at time $t = 0$. To include the effects of finite detector time resolution, we now convolve the time-dependent magnetic field strength with some smearing function, assumed here to be Gaussian with width $\sigma_t$, giving rise to the observed signal
\begin{align}\label{eq:Bint_exp}
    \vec{B}^\mathrm{obs}_\nu(t) &= \frac{1}{\sigma_t \sqrt{2\pi}} \int_{-\infty}^{\infty}\mathrm{d}t' \,  \vec{B}_\nu(t')\exp\left(-\frac{1}{2}\left[\frac{t'-t}{\sigma_t}\right]^2\right), \notag \\ 
    &= \vec{B}_{\nu}(t)\big|_{\sigma_\nu \to \widetilde\sigma_\nu} \\
    \widetilde\sigma_\nu &= \sqrt{\sigma_\nu^2 + (c\sigma_t)^2}.
\end{align}
Now we focus on the two regimes. When the spatial resolution, $\sigma_r = c \, \sigma_t$, is much smaller than the bunch size, $\sigma_r \ll \sigma_\nu$, only the $n = 0$ bunch gives a non-negligible contribution to the signal, which at $t = 0$ is given by
\begin{align}\label{eq:Bnu_good_res}
    B_{\nu,\max}^\mathrm{obs} &\simeq N_\nu \frac{q_\nu \, e}{(2\pi)^\frac{3}{2} \epsilon_0 c} \int \mathrm{d}^3p_\nu\, f_p(\vec{p}_{\nu})\frac{|\vec{\beta}_\nu|}{R_\perp \sigma_\nu} \notag \\
    & \simeq N_\nu \frac{q_\nu \, e}{(2\pi)^\frac{3}{2} \epsilon_0 c} \frac{1}{R_\perp \sigma_\nu},
\end{align}
to leading order in $R_\perp/\gamma_\nu \sigma_\nu$, an expansion that remains valid provided that the neutrinos are ultrarelativistic over the entire distribution function. The second line of~\eqref{eq:Bnu_good_res} similarly follows from assuming $|\vec{\beta}_\nu| = 1$ over the entire distribution function, and using the normalisation of $f_p$. From this it should be clear that provided the neutrino beam is ultrarelativistic, the shape of its momentum distribution function has no effect on the signal.

The other possibility is that spatial resolution is bad, encompassing several bunches such that $\sigma_r = 2n_r\Delta$. That is $\sigma_r \gg \sigma_\nu$, and $n\Delta$ for $|n| \ll n_r$. We can therefore truncate the sum for $|n| > n_r$, as these bunches will have a negligible contribution to the signal. Focusing on when one bunch is exactly centred at $t = 0$, and all neutrinos are ultrarelativistic we find
\begin{align}
    B_{\nu,\max}^\mathrm{obs} &\simeq N_\nu \frac{q_\nu \, e}{(2\pi)^\frac{3}{2} \epsilon_0 c} \sum_{n=-n_r}^{n_r}\frac{1}{R_\perp \sigma_r} \notag \\
    &= N_\nu \frac{ q_\nu \, e}{(2\pi)^\frac{3}{2} \epsilon_0 c} \frac{1}{R_\perp \Delta},
\end{align}
which is exactly the result~\eqref{eq:Bnu_good_res} with $\sigma_\nu \to \Delta$. One can think of this as diluting the effective neutrino density from $N_\nu/\sigma_\nu \to N_\nu/\Delta$, which is effectively what the detector sees.

\bibliography{bibliography}

\providecommand{\noopsort}[1]{}\providecommand{\singleletter}[1]{#1}%
\begin{thebibliography}{42}%
\makeatletter
\providecommand \@ifxundefined [1]{%
 \@ifx{#1\undefined}
}%
\providecommand \@ifnum [1]{%
 \ifnum #1\expandafter \@firstoftwo
 \else \expandafter \@secondoftwo
 \fi
}%
\providecommand \@ifx [1]{%
 \ifx #1\expandafter \@firstoftwo
 \else \expandafter \@secondoftwo
 \fi
}%
\providecommand \natexlab [1]{#1}%
\providecommand \enquote  [1]{``#1''}%
\providecommand \bibnamefont  [1]{#1}%
\providecommand \bibfnamefont [1]{#1}%
\providecommand \citenamefont [1]{#1}%
\providecommand \href@noop [0]{\@secondoftwo}%
\providecommand \href [0]{\begingroup \@sanitize@url \@href}%
\providecommand \@href[1]{\@@startlink{#1}\@@href}%
\providecommand \@@href[1]{\endgroup#1\@@endlink}%
\providecommand \@sanitize@url [0]{\catcode `\\12\catcode `\$12\catcode `\&12\catcode `\#12\catcode `\^12\catcode `\_12\catcode `\%12\relax}%
\providecommand \@@startlink[1]{}%
\providecommand \@@endlink[0]{}%
\providecommand \url  [0]{\begingroup\@sanitize@url \@url }%
\providecommand \@url [1]{\endgroup\@href {#1}{\urlprefix }}%
\providecommand \urlprefix  [0]{URL }%
\providecommand \Eprint [0]{\href }%
\providecommand \doibase [0]{https://doi.org/}%
\providecommand \selectlanguage [0]{\@gobble}%
\providecommand \bibinfo  [0]{\@secondoftwo}%
\providecommand \bibfield  [0]{\@secondoftwo}%
\providecommand \translation [1]{[#1]}%
\providecommand \BibitemOpen [0]{}%
\providecommand \bibitemStop [0]{}%
\providecommand \bibitemNoStop [0]{.\EOS\space}%
\providecommand \EOS [0]{\spacefactor3000\relax}%
\providecommand \BibitemShut  [1]{\csname bibitem#1\endcsname}%
\let\auto@bib@innerbib\@empty
\bibitem [{\citenamefont {Giunti}\ and\ \citenamefont {Studenikin}(2015)}]{Giunti:2014ixa}%
  \BibitemOpen
  \bibfield  {author} {\bibinfo {author} {\bibfnamefont {C.}~\bibnamefont {Giunti}}\ and\ \bibinfo {author} {\bibfnamefont {A.}~\bibnamefont {Studenikin}},\ }\href@noop {} {\bibfield  {journal} {\bibinfo  {journal} {Rev. Mod. Phys.}\ }\textbf {\bibinfo {volume} {87}},\ \bibinfo {pages} {531} (\bibinfo {year} {2015})},\ \Eprint {https://arxiv.org/abs/1403.6344} {arXiv:1403.6344 [hep-ph]} \BibitemShut {NoStop}%
\bibitem [{\citenamefont {Babu}\ and\ \citenamefont {Mohapatra}(1989)}]{Babu:1989tq}%
  \BibitemOpen
  \bibfield  {author} {\bibinfo {author} {\bibfnamefont {K.~S.}\ \bibnamefont {Babu}}\ and\ \bibinfo {author} {\bibfnamefont {R.~N.}\ \bibnamefont {Mohapatra}},\ }\href@noop {} {\bibfield  {journal} {\bibinfo  {journal} {Phys. Rev. Lett.}\ }\textbf {\bibinfo {volume} {63}},\ \bibinfo {pages} {938} (\bibinfo {year} {1989})}\BibitemShut {NoStop}%
\bibitem [{\citenamefont {Babu}\ and\ \citenamefont {Mohapatra}(1990)}]{Babu:1989ex}%
  \BibitemOpen
  \bibfield  {author} {\bibinfo {author} {\bibfnamefont {K.~S.}\ \bibnamefont {Babu}}\ and\ \bibinfo {author} {\bibfnamefont {R.~N.}\ \bibnamefont {Mohapatra}},\ }\href@noop {} {\bibfield  {journal} {\bibinfo  {journal} {Phys. Rev. D}\ }\textbf {\bibinfo {volume} {41}},\ \bibinfo {pages} {271} (\bibinfo {year} {1990})}\BibitemShut {NoStop}%
\bibitem [{\citenamefont {Foot}\ \emph {et~al.}(1990)\citenamefont {Foot}, \citenamefont {Joshi}, \citenamefont {Lew},\ and\ \citenamefont {Volkas}}]{Foot:1990uf}%
  \BibitemOpen
  \bibfield  {author} {\bibinfo {author} {\bibfnamefont {R.}~\bibnamefont {Foot}}, \bibinfo {author} {\bibfnamefont {G.~C.}\ \bibnamefont {Joshi}}, \bibinfo {author} {\bibfnamefont {H.}~\bibnamefont {Lew}},\ and\ \bibinfo {author} {\bibfnamefont {R.~R.}\ \bibnamefont {Volkas}},\ }\href@noop {} {\bibfield  {journal} {\bibinfo  {journal} {Mod. Phys. Lett. A}\ }\textbf {\bibinfo {volume} {5}},\ \bibinfo {pages} {2721} (\bibinfo {year} {1990})}\BibitemShut {NoStop}%
\bibitem [{\citenamefont {Foot}\ \emph {et~al.}(1993)\citenamefont {Foot}, \citenamefont {Lew},\ and\ \citenamefont {Volkas}}]{Foot:1992ui}%
  \BibitemOpen
  \bibfield  {author} {\bibinfo {author} {\bibfnamefont {R.}~\bibnamefont {Foot}}, \bibinfo {author} {\bibfnamefont {H.}~\bibnamefont {Lew}},\ and\ \bibinfo {author} {\bibfnamefont {R.~R.}\ \bibnamefont {Volkas}},\ }\href@noop {} {\bibfield  {journal} {\bibinfo  {journal} {J. Phys. G}\ }\textbf {\bibinfo {volume} {19}},\ \bibinfo {pages} {361} (\bibinfo {year} {1993})},\ \bibinfo {note} {[Erratum: J.Phys.G 19, 1067 (1993)]},\ \Eprint {https://arxiv.org/abs/hep-ph/9209259} {arXiv:hep-ph/9209259} \BibitemShut {NoStop}%
\bibitem [{\citenamefont {Giunti}\ \emph {et~al.}(2024)\citenamefont {Giunti}, \citenamefont {Kouzakov}, \citenamefont {Li},\ and\ \citenamefont {Studenikin}}]{Giunti:2024gec}%
  \BibitemOpen
  \bibfield  {author} {\bibinfo {author} {\bibfnamefont {C.}~\bibnamefont {Giunti}}, \bibinfo {author} {\bibfnamefont {K.}~\bibnamefont {Kouzakov}}, \bibinfo {author} {\bibfnamefont {Y.-F.}\ \bibnamefont {Li}},\ and\ \bibinfo {author} {\bibfnamefont {A.}~\bibnamefont {Studenikin}},\ }\href@noop {} {} (\bibinfo {year} {2024}),\ \Eprint {https://arxiv.org/abs/2411.03122} {arXiv:2411.03122 [hep-ph]} \BibitemShut {NoStop}%
\bibitem [{\citenamefont {Caprini}\ \emph {et~al.}(2005)\citenamefont {Caprini}, \citenamefont {Biller},\ and\ \citenamefont {Ferreira}}]{Caprini:2003gz}%
  \BibitemOpen
  \bibfield  {author} {\bibinfo {author} {\bibfnamefont {C.}~\bibnamefont {Caprini}}, \bibinfo {author} {\bibfnamefont {S.}~\bibnamefont {Biller}},\ and\ \bibinfo {author} {\bibfnamefont {P.~G.}\ \bibnamefont {Ferreira}},\ }\href@noop {} {\bibfield  {journal} {\bibinfo  {journal} {JCAP}\ }\textbf {\bibinfo {volume} {02}},\ \bibinfo {pages} {006}},\ \Eprint {https://arxiv.org/abs/hep-ph/0310066} {arXiv:hep-ph/0310066} \BibitemShut {NoStop}%
\bibitem [{\citenamefont {Navas}\ \emph {et~al.}(2024)\citenamefont {Navas} \emph {et~al.}}]{ParticleDataGroup:2024cfk}%
  \BibitemOpen
  \bibfield  {author} {\bibinfo {author} {\bibfnamefont {S.}~\bibnamefont {Navas}} \emph {et~al.} (\bibinfo {collaboration} {Particle Data Group}),\ }\href@noop {} {\bibfield  {journal} {\bibinfo  {journal} {Phys. Rev. D}\ }\textbf {\bibinfo {volume} {110}},\ \bibinfo {pages} {030001} (\bibinfo {year} {2024})}\BibitemShut {NoStop}%
\bibitem [{\citenamefont {Studenikin}\ and\ \citenamefont {Tokarev}(2014)}]{Studenikin:2012vi}%
  \BibitemOpen
  \bibfield  {author} {\bibinfo {author} {\bibfnamefont {A.~I.}\ \bibnamefont {Studenikin}}\ and\ \bibinfo {author} {\bibfnamefont {I.}~\bibnamefont {Tokarev}},\ }\href@noop {} {\bibfield  {journal} {\bibinfo  {journal} {Nucl. Phys. B}\ }\textbf {\bibinfo {volume} {884}},\ \bibinfo {pages} {396} (\bibinfo {year} {2014})},\ \Eprint {https://arxiv.org/abs/1209.3245} {arXiv:1209.3245 [hep-ph]} \BibitemShut {NoStop}%
\bibitem [{\citenamefont {Atzori~Corona}\ \emph {et~al.}(2023)\citenamefont {Atzori~Corona}, \citenamefont {Bonivento}, \citenamefont {Cadeddu}, \citenamefont {Cargioli},\ and\ \citenamefont {Dordei}}]{AtzoriCorona:2022jeb}%
  \BibitemOpen
  \bibfield  {author} {\bibinfo {author} {\bibfnamefont {M.}~\bibnamefont {Atzori~Corona}}, \bibinfo {author} {\bibfnamefont {W.~M.}\ \bibnamefont {Bonivento}}, \bibinfo {author} {\bibfnamefont {M.}~\bibnamefont {Cadeddu}}, \bibinfo {author} {\bibfnamefont {N.}~\bibnamefont {Cargioli}},\ and\ \bibinfo {author} {\bibfnamefont {F.}~\bibnamefont {Dordei}},\ }\href@noop {} {\bibfield  {journal} {\bibinfo  {journal} {Phys. Rev. D}\ }\textbf {\bibinfo {volume} {107}},\ \bibinfo {pages} {053001} (\bibinfo {year} {2023})},\ \Eprint {https://arxiv.org/abs/2207.05036} {arXiv:2207.05036 [hep-ph]} \BibitemShut {NoStop}%
\bibitem [{\citenamefont {ShivaSankar}\ \emph {et~al.}(2023)\citenamefont {ShivaSankar}, \citenamefont {Majumdar}, \citenamefont {Papoulias}, \citenamefont {Prajapati},\ and\ \citenamefont {Srivastava}}]{A:2022acy}%
  \BibitemOpen
  \bibfield  {author} {\bibinfo {author} {\bibfnamefont {K.~A.}\ \bibnamefont {ShivaSankar}}, \bibinfo {author} {\bibfnamefont {A.}~\bibnamefont {Majumdar}}, \bibinfo {author} {\bibfnamefont {D.~K.}\ \bibnamefont {Papoulias}}, \bibinfo {author} {\bibfnamefont {H.}~\bibnamefont {Prajapati}},\ and\ \bibinfo {author} {\bibfnamefont {R.}~\bibnamefont {Srivastava}},\ }\href {https://doi.org/10.1016/j.physletb.2023.137742} {\bibfield  {journal} {\bibinfo  {journal} {Phys. Lett. B}\ }\textbf {\bibinfo {volume} {839}},\ \bibinfo {pages} {137742} (\bibinfo {year} {2023})},\ \Eprint {https://arxiv.org/abs/2208.06415} {arXiv:2208.06415 [hep-ph]} \BibitemShut {NoStop}%
\bibitem [{\citenamefont {Giunti}\ and\ \citenamefont {Ternes}(2023)}]{Giunti:2023yha}%
  \BibitemOpen
  \bibfield  {author} {\bibinfo {author} {\bibfnamefont {C.}~\bibnamefont {Giunti}}\ and\ \bibinfo {author} {\bibfnamefont {C.~A.}\ \bibnamefont {Ternes}},\ }\href@noop {} {\bibfield  {journal} {\bibinfo  {journal} {Phys. Rev. D}\ }\textbf {\bibinfo {volume} {108}},\ \bibinfo {pages} {095044} (\bibinfo {year} {2023})},\ \Eprint {https://arxiv.org/abs/2309.17380} {arXiv:2309.17380 [hep-ph]} \BibitemShut {NoStop}%
\bibitem [{\citenamefont {Jackson}(1998)}]{Jackson:1998nia}%
  \BibitemOpen
  \bibfield  {author} {\bibinfo {author} {\bibfnamefont {J.~D.}\ \bibnamefont {Jackson}},\ }\href@noop {} {\emph {\bibinfo {title} {{Classical Electrodynamics}}}}\ (\bibinfo  {publisher} {Wiley},\ \bibinfo {year} {1998})\BibitemShut {NoStop}%
\bibitem [{\citenamefont {Acciarri}\ \emph {et~al.}(2015)\citenamefont {Acciarri} \emph {et~al.}}]{DUNE:2015lol}%
  \BibitemOpen
  \bibfield  {author} {\bibinfo {author} {\bibfnamefont {R.}~\bibnamefont {Acciarri}} \emph {et~al.} (\bibinfo {collaboration} {DUNE}),\ }\href@noop {} {\  (\bibinfo {year} {2015})},\ \Eprint {https://arxiv.org/abs/1512.06148} {arXiv:1512.06148 [physics.ins-det]} \BibitemShut {NoStop}%
\bibitem [{JPA(2003)}]{JPARC}%
  \BibitemOpen
  \href@noop {} {\emph {\bibinfo {title} {{Accelerator technical design report for high-intensity proton accelerator facility project, J-PARC}}}},\ \bibinfo {type} {Tech. Rep.}\ \bibinfo {number} {JAERI-TECH-2003-044, KEK-2002-13}\ (\bibinfo  {institution} {JAERI, KEK},\ \bibinfo {year} {2003})\BibitemShut {NoStop}%
\bibitem [{\citenamefont {Jackson~Kimball}\ \emph {et~al.}(2020)\citenamefont {Jackson~Kimball} \emph {et~al.}}]{JacksonKimball:2017elr}%
  \BibitemOpen
  \bibfield  {author} {\bibinfo {author} {\bibfnamefont {D.~F.}\ \bibnamefont {Jackson~Kimball}} \emph {et~al.},\ }\href@noop {} {\bibfield  {journal} {\bibinfo  {journal} {Springer Proc. Phys.}\ }\textbf {\bibinfo {volume} {245}},\ \bibinfo {pages} {105} (\bibinfo {year} {2020})},\ \Eprint {https://arxiv.org/abs/1711.08999} {arXiv:1711.08999 [physics.ins-det]} \BibitemShut {NoStop}%
\bibitem [{\citenamefont {Kominis}\ \emph {et~al.}(2003)\citenamefont {Kominis}, \citenamefont {Kornack}, \citenamefont {Allred},\ and\ \citenamefont {Romalis}}]{Kominis:2003gax}%
  \BibitemOpen
  \bibfield  {author} {\bibinfo {author} {\bibfnamefont {I.~K.}\ \bibnamefont {Kominis}}, \bibinfo {author} {\bibfnamefont {T.~W.}\ \bibnamefont {Kornack}}, \bibinfo {author} {\bibfnamefont {J.~C.}\ \bibnamefont {Allred}},\ and\ \bibinfo {author} {\bibfnamefont {M.~V.}\ \bibnamefont {Romalis}},\ }\href@noop {} {\bibfield  {journal} {\bibinfo  {journal} {Nature}\ }\textbf {\bibinfo {volume} {422}},\ \bibinfo {pages} {596} (\bibinfo {year} {2003})}\BibitemShut {NoStop}%
\bibitem [{\citenamefont {Ganguly}\ \emph {et~al.}(2024)\citenamefont {Ganguly}, \citenamefont {Yonehara}, \citenamefont {Bhat}, \citenamefont {Triplett}, \citenamefont {Ainsworth}, \citenamefont {Hinds},\ and\ \citenamefont {Abdelhamid}}]{Ganguly:2024lqh}%
  \BibitemOpen
  \bibfield  {author} {\bibinfo {author} {\bibfnamefont {S.}~\bibnamefont {Ganguly}}, \bibinfo {author} {\bibfnamefont {K.}~\bibnamefont {Yonehara}}, \bibinfo {author} {\bibfnamefont {C.~M.}\ \bibnamefont {Bhat}}, \bibinfo {author} {\bibfnamefont {A.~K.}\ \bibnamefont {Triplett}}, \bibinfo {author} {\bibfnamefont {R.}~\bibnamefont {Ainsworth}}, \bibinfo {author} {\bibfnamefont {C.}~\bibnamefont {Hinds}},\ and\ \bibinfo {author} {\bibfnamefont {M.}~\bibnamefont {Abdelhamid}},\ }\href@noop {} {\  (\bibinfo {year} {2024})},\ \Eprint {https://arxiv.org/abs/2410.18256} {arXiv:2410.18256 [physics.acc-ph]} \BibitemShut {NoStop}%
\bibitem [{\citenamefont {Strait}\ \emph {et~al.}(2016)\citenamefont {Strait} \emph {et~al.}}]{DUNE:2016evb}%
  \BibitemOpen
  \bibfield  {author} {\bibinfo {author} {\bibfnamefont {J.}~\bibnamefont {Strait}} \emph {et~al.} (\bibinfo {collaboration} {DUNE}),\ }\href@noop {} {} (\bibinfo {year} {2016}),\ \Eprint {https://arxiv.org/abs/1601.05823} {arXiv:1601.05823 [physics.ins-det]} \BibitemShut {NoStop}%
\bibitem [{\citenamefont {Angelico}\ \emph {et~al.}(2019)\citenamefont {Angelico}, \citenamefont {Eisch}, \citenamefont {Elagin}, \citenamefont {Frisch}, \citenamefont {Nagaitsev},\ and\ \citenamefont {Wetstein}}]{Angelico:2019gyi}%
  \BibitemOpen
  \bibfield  {author} {\bibinfo {author} {\bibfnamefont {E.}~\bibnamefont {Angelico}}, \bibinfo {author} {\bibfnamefont {J.}~\bibnamefont {Eisch}}, \bibinfo {author} {\bibfnamefont {A.}~\bibnamefont {Elagin}}, \bibinfo {author} {\bibfnamefont {H.}~\bibnamefont {Frisch}}, \bibinfo {author} {\bibfnamefont {S.}~\bibnamefont {Nagaitsev}},\ and\ \bibinfo {author} {\bibfnamefont {M.}~\bibnamefont {Wetstein}},\ }\href@noop {} {\bibfield  {journal} {\bibinfo  {journal} {Phys. Rev. D}\ }\textbf {\bibinfo {volume} {100}},\ \bibinfo {pages} {032008} (\bibinfo {year} {2019})},\ \Eprint {https://arxiv.org/abs/1904.01611} {arXiv:1904.01611 [physics.acc-ph]} \BibitemShut {NoStop}%
\bibitem [{\citenamefont {Facon}\ \emph {et~al.}(2016)\citenamefont {Facon}, \citenamefont {Dietsche}, \citenamefont {Grosso}, \citenamefont {Haroche}, \citenamefont {Raimond}, \citenamefont {Brune},\ and\ \citenamefont {Gleyzes}}]{Facon:2016ary}%
  \BibitemOpen
  \bibfield  {author} {\bibinfo {author} {\bibfnamefont {A.}~\bibnamefont {Facon}}, \bibinfo {author} {\bibfnamefont {E.-K.}\ \bibnamefont {Dietsche}}, \bibinfo {author} {\bibfnamefont {D.}~\bibnamefont {Grosso}}, \bibinfo {author} {\bibfnamefont {S.}~\bibnamefont {Haroche}}, \bibinfo {author} {\bibfnamefont {J.-M.}\ \bibnamefont {Raimond}}, \bibinfo {author} {\bibfnamefont {M.}~\bibnamefont {Brune}},\ and\ \bibinfo {author} {\bibfnamefont {S.}~\bibnamefont {Gleyzes}},\ }\href@noop {} {\bibfield  {journal} {\bibinfo  {journal} {Nature}\ }\textbf {\bibinfo {volume} {535}},\ \bibinfo {pages} {262} (\bibinfo {year} {2016})}\BibitemShut {NoStop}%
\bibitem [{\citenamefont {Friend}(2019)}]{Friend:2019fuq}%
  \BibitemOpen
  \bibfield  {author} {\bibinfo {author} {\bibfnamefont {M.}~\bibnamefont {Friend}},\ }in\ \href@noop {} {\emph {\bibinfo {booktitle} {Proc. 7th International Beam Instrumentation Conference (IBIC'18), Shanghai, China, 09-13 September 2018}}}\ (\bibinfo  {publisher} {JACoW Publishing},\ \bibinfo {address} {Geneva, Switzerland},\ \bibinfo {year} {2019})\ pp.\ \bibinfo {pages} {85--88}\BibitemShut {NoStop}%
\bibitem [{\citenamefont {Aguilar-Arevalo}\ \emph {et~al.}(2022)\citenamefont {Aguilar-Arevalo} \emph {et~al.}}]{CCM:2021leg}%
  \BibitemOpen
  \bibfield  {author} {\bibinfo {author} {\bibfnamefont {A.~A.}\ \bibnamefont {Aguilar-Arevalo}} \emph {et~al.} (\bibinfo {collaboration} {CCM}),\ }\href@noop {} {\bibfield  {journal} {\bibinfo  {journal} {Phys. Rev. D}\ }\textbf {\bibinfo {volume} {106}},\ \bibinfo {pages} {012001} (\bibinfo {year} {2022})},\ \Eprint {https://arxiv.org/abs/2105.14020} {arXiv:2105.14020 [hep-ex]} \BibitemShut {NoStop}%
\bibitem [{\citenamefont {Alekou}\ \emph {et~al.}(2023)\citenamefont {Alekou} \emph {et~al.}}]{ESSnuSB:2023ogw}%
  \BibitemOpen
  \bibfield  {author} {\bibinfo {author} {\bibfnamefont {A.}~\bibnamefont {Alekou}} \emph {et~al.} (\bibinfo {collaboration} {ESSnuSB}),\ }\href@noop {} {\bibfield  {journal} {\bibinfo  {journal} {Universe}\ }\textbf {\bibinfo {volume} {9}},\ \bibinfo {pages} {347} (\bibinfo {year} {2023})},\ \Eprint {https://arxiv.org/abs/2303.17356} {arXiv:2303.17356 [hep-ex]} \BibitemShut {NoStop}%
\bibitem [{\citenamefont {Garoby}\ \emph {et~al.}(2018)\citenamefont {Garoby} \emph {et~al.}}]{Garoby:2017vew}%
  \BibitemOpen
  \bibfield  {author} {\bibinfo {author} {\bibfnamefont {R.}~\bibnamefont {Garoby}} \emph {et~al.},\ }\href@noop {} {\bibfield  {journal} {\bibinfo  {journal} {Phys. Scripta}\ }\textbf {\bibinfo {volume} {93}},\ \bibinfo {pages} {014001} (\bibinfo {year} {2018})}\BibitemShut {NoStop}%
\bibitem [{\citenamefont {Ahdida}\ \emph {et~al.}(2020)\citenamefont {Ahdida} \emph {et~al.}}]{Ahdida:2020whw}%
  \BibitemOpen
  \bibfield  {author} {\bibinfo {author} {\bibfnamefont {C.}~\bibnamefont {Ahdida}} \emph {et~al.},\ }\href@noop {} {\emph {\bibinfo {title} {{nuSTORM at CERN: Feasibility Study}}}},\ \bibinfo {type} {Tech. Rep.}\ (\bibinfo  {institution} {CERN},\ \bibinfo {address} {Geneva},\ \bibinfo {year} {2020})\BibitemShut {NoStop}%
\bibitem [{\citenamefont {Longhin}\ and\ \citenamefont {Terranova}(2020)}]{Longhin:2714046}%
  \BibitemOpen
  \bibfield  {author} {\bibinfo {author} {\bibfnamefont {A.}~\bibnamefont {Longhin}}\ and\ \bibinfo {author} {\bibfnamefont {F.}~\bibnamefont {Terranova}},\ }\href@noop {} {\emph {\bibinfo {title} {{NP06/ENUBET annual report for the CERN-SPSC}}}},\ \bibinfo {type} {Tech. Rep.}\ (\bibinfo  {institution} {CERN},\ \bibinfo {address} {Geneva},\ \bibinfo {year} {2020})\BibitemShut {NoStop}%
\bibitem [{\citenamefont {Acerbi}\ \emph {et~al.}(2023)\citenamefont {Acerbi} \emph {et~al.}}]{ENUBET:2023hgu}%
  \BibitemOpen
  \bibfield  {author} {\bibinfo {author} {\bibfnamefont {F.}~\bibnamefont {Acerbi}} \emph {et~al.} (\bibinfo {collaboration} {ENUBET}),\ }\href@noop {} {\bibfield  {journal} {\bibinfo  {journal} {Eur. Phys. J. C}\ }\textbf {\bibinfo {volume} {83}},\ \bibinfo {pages} {964} (\bibinfo {year} {2023})},\ \Eprint {https://arxiv.org/abs/2308.09402} {arXiv:2308.09402 [hep-ex]} \BibitemShut {NoStop}%
\bibitem [{\citenamefont {Aprili}(2006)}]{Aprili:2006}%
  \BibitemOpen
  \bibfield  {author} {\bibinfo {author} {\bibfnamefont {M.}~\bibnamefont {Aprili}},\ }\href {https://doi.org/10.1038/nnano.2006.78} {\bibfield  {journal} {\bibinfo  {journal} {Nature Nanotechnology}\ }\textbf {\bibinfo {volume} {1}},\ \bibinfo {pages} {15} (\bibinfo {year} {2006})}\BibitemShut {NoStop}%
\bibitem [{\citenamefont {Bauer}\ \emph {et~al.}(2018)\citenamefont {Bauer}, \citenamefont {Foldenauer},\ and\ \citenamefont {Jaeckel}}]{Bauer:2018onh}%
  \BibitemOpen
  \bibfield  {author} {\bibinfo {author} {\bibfnamefont {M.}~\bibnamefont {Bauer}}, \bibinfo {author} {\bibfnamefont {P.}~\bibnamefont {Foldenauer}},\ and\ \bibinfo {author} {\bibfnamefont {J.}~\bibnamefont {Jaeckel}},\ }\href@noop {} {\bibfield  {journal} {\bibinfo  {journal} {JHEP}\ }\textbf {\bibinfo {volume} {07}},\ \bibinfo {pages} {094}},\ \Eprint {https://arxiv.org/abs/1803.05466} {arXiv:1803.05466 [hep-ph]} \BibitemShut {NoStop}%
\bibitem [{\citenamefont {Cline}(2024)}]{Cline:2024qzv}%
  \BibitemOpen
  \bibfield  {author} {\bibinfo {author} {\bibfnamefont {J.~M.}\ \bibnamefont {Cline}},\ }in\ \href@noop {} {\emph {\bibinfo {booktitle} {{58th Rencontres de Moriond on Electroweak Interactions and Unified Theories}}}}\ (\bibinfo {year} {2024})\BibitemShut {NoStop}%
\bibitem [{\citenamefont {Foldenauer}(2019)}]{Foldenauer:2019vgn}%
  \BibitemOpen
  \bibfield  {author} {\bibinfo {author} {\bibfnamefont {P.}~\bibnamefont {Foldenauer}},\ }\emph {\bibinfo {title} {{Phenomenology of Extra Abelian Gauge Symmetries}}},\ \href@noop {} {Ph.D. thesis},\ \bibinfo  {school} {U. Heidelberg (main)} (\bibinfo {year} {2019})\BibitemShut {NoStop}%
\bibitem [{\citenamefont {Wise}\ and\ \citenamefont {Zhang}(2018)}]{Wise:2018rnb}%
  \BibitemOpen
  \bibfield  {author} {\bibinfo {author} {\bibfnamefont {M.~B.}\ \bibnamefont {Wise}}\ and\ \bibinfo {author} {\bibfnamefont {Y.}~\bibnamefont {Zhang}},\ }\href {https://doi.org/10.1007/JHEP06(2018)053} {\bibfield  {journal} {\bibinfo  {journal} {JHEP}\ }\textbf {\bibinfo {volume} {06}},\ \bibinfo {pages} {053}},\ \Eprint {https://arxiv.org/abs/1803.00591} {arXiv:1803.00591 [hep-ph]} \BibitemShut {NoStop}%
\bibitem [{\citenamefont {Adelberger}\ \emph {et~al.}(2009)\citenamefont {Adelberger}, \citenamefont {Gundlach}, \citenamefont {Heckel}, \citenamefont {Hoedl},\ and\ \citenamefont {Schlamminger}}]{Adelberger:2009zz}%
  \BibitemOpen
  \bibfield  {author} {\bibinfo {author} {\bibfnamefont {E.~G.}\ \bibnamefont {Adelberger}}, \bibinfo {author} {\bibfnamefont {J.~H.}\ \bibnamefont {Gundlach}}, \bibinfo {author} {\bibfnamefont {B.~R.}\ \bibnamefont {Heckel}}, \bibinfo {author} {\bibfnamefont {S.}~\bibnamefont {Hoedl}},\ and\ \bibinfo {author} {\bibfnamefont {S.}~\bibnamefont {Schlamminger}},\ }\href {https://doi.org/10.1016/j.ppnp.2008.08.002} {\bibfield  {journal} {\bibinfo  {journal} {Prog. Part. Nucl. Phys.}\ }\textbf {\bibinfo {volume} {62}},\ \bibinfo {pages} {102} (\bibinfo {year} {2009})}\BibitemShut {NoStop}%
\bibitem [{\citenamefont {Wagner}\ \emph {et~al.}(2012)\citenamefont {Wagner}, \citenamefont {Schlamminger}, \citenamefont {Gundlach},\ and\ \citenamefont {Adelberger}}]{Wagner:2012ui}%
  \BibitemOpen
  \bibfield  {author} {\bibinfo {author} {\bibfnamefont {T.~A.}\ \bibnamefont {Wagner}}, \bibinfo {author} {\bibfnamefont {S.}~\bibnamefont {Schlamminger}}, \bibinfo {author} {\bibfnamefont {J.~H.}\ \bibnamefont {Gundlach}},\ and\ \bibinfo {author} {\bibfnamefont {E.~G.}\ \bibnamefont {Adelberger}},\ }\href {https://doi.org/10.1088/0264-9381/29/18/184002} {\bibfield  {journal} {\bibinfo  {journal} {Class. Quant. Grav.}\ }\textbf {\bibinfo {volume} {29}},\ \bibinfo {pages} {184002} (\bibinfo {year} {2012})},\ \Eprint {https://arxiv.org/abs/1207.2442} {arXiv:1207.2442 [gr-qc]} \BibitemShut {NoStop}%
\bibitem [{\citenamefont {Venzor}\ \emph {et~al.}(2021)\citenamefont {Venzor}, \citenamefont {P\'erez-Lorenzana},\ and\ \citenamefont {De-Santiago}}]{Venzor:2020ova}%
  \BibitemOpen
  \bibfield  {author} {\bibinfo {author} {\bibfnamefont {J.}~\bibnamefont {Venzor}}, \bibinfo {author} {\bibfnamefont {A.}~\bibnamefont {P\'erez-Lorenzana}},\ and\ \bibinfo {author} {\bibfnamefont {J.}~\bibnamefont {De-Santiago}},\ }\href {https://doi.org/10.1103/PhysRevD.103.043534} {\bibfield  {journal} {\bibinfo  {journal} {Phys. Rev. D}\ }\textbf {\bibinfo {volume} {103}},\ \bibinfo {pages} {043534} (\bibinfo {year} {2021})},\ \Eprint {https://arxiv.org/abs/2009.08104} {arXiv:2009.08104 [hep-ph]} \BibitemShut {NoStop}%
\bibitem [{\citenamefont {Caputo}\ \emph {et~al.}(2020)\citenamefont {Caputo}, \citenamefont {Liu}, \citenamefont {Mishra-Sharma},\ and\ \citenamefont {Ruderman}}]{Caputo:2020bdy}%
  \BibitemOpen
  \bibfield  {author} {\bibinfo {author} {\bibfnamefont {A.}~\bibnamefont {Caputo}}, \bibinfo {author} {\bibfnamefont {H.}~\bibnamefont {Liu}}, \bibinfo {author} {\bibfnamefont {S.}~\bibnamefont {Mishra-Sharma}},\ and\ \bibinfo {author} {\bibfnamefont {J.~T.}\ \bibnamefont {Ruderman}},\ }\href@noop {} {\bibfield  {journal} {\bibinfo  {journal} {Phys. Rev. Lett.}\ }\textbf {\bibinfo {volume} {125}},\ \bibinfo {pages} {221303} (\bibinfo {year} {2020})},\ \Eprint {https://arxiv.org/abs/2002.05165} {arXiv:2002.05165 [astro-ph.CO]} \BibitemShut {NoStop}%
\bibitem [{\citenamefont {Garcia}\ \emph {et~al.}(2020)\citenamefont {Garcia}, \citenamefont {Bondarenko}, \citenamefont {Ploeckinger}, \citenamefont {Pradler},\ and\ \citenamefont {Sokolenko}}]{Garcia:2020qrp}%
  \BibitemOpen
  \bibfield  {author} {\bibinfo {author} {\bibfnamefont {A.~A.}\ \bibnamefont {Garcia}}, \bibinfo {author} {\bibfnamefont {K.}~\bibnamefont {Bondarenko}}, \bibinfo {author} {\bibfnamefont {S.}~\bibnamefont {Ploeckinger}}, \bibinfo {author} {\bibfnamefont {J.}~\bibnamefont {Pradler}},\ and\ \bibinfo {author} {\bibfnamefont {A.}~\bibnamefont {Sokolenko}},\ }\href@noop {} {\bibfield  {journal} {\bibinfo  {journal} {JCAP}\ }\textbf {\bibinfo {volume} {10}},\ \bibinfo {pages} {011}},\ \Eprint {https://arxiv.org/abs/2003.10465} {arXiv:2003.10465 [astro-ph.CO]} \BibitemShut {NoStop}%
\bibitem [{\citenamefont {Betz}\ \emph {et~al.}(2013)\citenamefont {Betz}, \citenamefont {Caspers}, \citenamefont {Gasior}, \citenamefont {Thumm},\ and\ \citenamefont {Rieger}}]{Betz:2013dza}%
  \BibitemOpen
  \bibfield  {author} {\bibinfo {author} {\bibfnamefont {M.}~\bibnamefont {Betz}}, \bibinfo {author} {\bibfnamefont {F.}~\bibnamefont {Caspers}}, \bibinfo {author} {\bibfnamefont {M.}~\bibnamefont {Gasior}}, \bibinfo {author} {\bibfnamefont {M.}~\bibnamefont {Thumm}},\ and\ \bibinfo {author} {\bibfnamefont {S.~W.}\ \bibnamefont {Rieger}},\ }\href@noop {} {\bibfield  {journal} {\bibinfo  {journal} {Phys. Rev. D}\ }\textbf {\bibinfo {volume} {88}},\ \bibinfo {pages} {075014} (\bibinfo {year} {2013})},\ \Eprint {https://arxiv.org/abs/1310.8098} {arXiv:1310.8098 [physics.ins-det]} \BibitemShut {NoStop}%
\bibitem [{\citenamefont {Jaeckel}\ and\ \citenamefont {Roy}(2010)}]{Jaeckel:2010xx}%
  \BibitemOpen
  \bibfield  {author} {\bibinfo {author} {\bibfnamefont {J.}~\bibnamefont {Jaeckel}}\ and\ \bibinfo {author} {\bibfnamefont {S.}~\bibnamefont {Roy}},\ }\href@noop {} {\bibfield  {journal} {\bibinfo  {journal} {Phys. Rev. D}\ }\textbf {\bibinfo {volume} {82}},\ \bibinfo {pages} {125020} (\bibinfo {year} {2010})},\ \Eprint {https://arxiv.org/abs/1008.3536} {arXiv:1008.3536 [hep-ph]} \BibitemShut {NoStop}%
\bibitem [{\citenamefont {Fabbrichesi}\ \emph {et~al.}(2021)\citenamefont {Fabbrichesi}, \citenamefont {Gabrielli},\ and\ \citenamefont {Lanfranchi}}]{Fabbrichesi2021}%
  \BibitemOpen
  \bibfield  {author} {\bibinfo {author} {\bibfnamefont {M.}~\bibnamefont {Fabbrichesi}}, \bibinfo {author} {\bibfnamefont {E.}~\bibnamefont {Gabrielli}},\ and\ \bibinfo {author} {\bibfnamefont {G.}~\bibnamefont {Lanfranchi}},\ }\href@noop {} {\emph {\bibinfo {title} {The Physics of the Dark Photon: A Primer}}}\ (\bibinfo  {publisher} {Springer International Publishing},\ \bibinfo {year} {2021})\BibitemShut {NoStop}%
\bibitem [{\citenamefont {Li}\ and\ \citenamefont {Xu}(2023)}]{Li:2023vpv}%
  \BibitemOpen
  \bibfield  {author} {\bibinfo {author} {\bibfnamefont {S.-P.}\ \bibnamefont {Li}}\ and\ \bibinfo {author} {\bibfnamefont {X.-J.}\ \bibnamefont {Xu}},\ }\href {https://doi.org/10.1088/1475-7516/2023/09/009} {\bibfield  {journal} {\bibinfo  {journal} {JCAP}\ }\textbf {\bibinfo {volume} {09}},\ \bibinfo {pages} {009}},\ \Eprint {https://arxiv.org/abs/2304.12907} {arXiv:2304.12907 [hep-ph]} \BibitemShut {NoStop}%
\end{thebibliography}%

\end{document}